# Demystifying Double Robustness: A Comparison of Alternative Strategies for Estimating a Population Mean from Incomplete Data[1]

**Joseph D. Y. Kang and Joseph L. Schafer**

*Abstract.* When outcomes are missing for reasons beyond an investigator's control, there are two different ways to adjust a parameter estimate for covariates that may be related both to the outcome and to missingness. One approach is to model the relationships between the covariates and the outcome and use those relationships to predict the missing values. Another is to model the probabilities of missingness given the covariates and incorporate them into a weighted or stratified estimate. Doubly robust (DR) procedures apply both types of model simultaneously and produce a consistent estimate of the parameter if either of the two models has been correctly specified. In this article, we show that DR estimates can be constructed in many ways. We compare the performance of various DR and non-DR estimates of a population mean in a simulated example where both models are incorrect but neither is grossly misspecified. Methods that use inverse-probabilities as weights, whether they are DR or not, are sensitive to misspecification of the propensity model when some estimated propensities are small. Many DR methods perform better than simple inverse-probability weighting. None of the DR methods we tried, however, improved upon the performance of simple regression-based prediction of the missing values. This study does not represent every missing-data problem that will arise in practice. But it does demonstrate that, in at least some settings, two wrong models are not better than one.

*Joseph D. Y. Kang is Research Associate, The Methodology Center, 204 E. Calder Way, Suite 400, State College, Pennsylvania 16801, USA e-mail: josephkang@stat.psu.edu. Joseph L. Schafer is Associate Professor, The Methodology Center, 204 E. Calder Way, Suite 400, State College, Pennsylvania 16801, USA e-mail: jls@stat.psu.edu.*
[1]Discussed in 10.1214/07-STS227A, 10.1214/07-STS227B, 10.1214/07-STS227C and 10.1214/07-STS227D; rejoinder at 10.1214/07-STS227REJ.

## 1. INTRODUCTION

### 1.1 Purpose

A new class of methods called doubly robust (DR) procedures was designed to mitigate selection bias arising from uncontrolled nonresponse and attrition, nonrandom treatment assignment in observational studies and noncompliance in randomized experiments (Robins and Rotnitzky, 2001; van der Laan and Robins, 2003). DR methods require specification of two models: one that describes the population of responses, and another that describes the process by which the data are filtered or selected to produce the observed sample. The distinguishing feature of DR estimates is that they remain asymptotically unbiased if one of the two models is misspecified—that is, they consistently estimate their targets if either model is true (Robins and Rotnitzky, 1995).

DR methods are a refinement of a weighted estimating-equations approach to regression with incomplete data proposed by Robins, Rotnitzky and Zhao (1994, 1995) and Rotnitzky, Robins and Scharfstein (1998). Further explanation and evaluation of DR estimators has been given by Lunceford and Davidian (2004), Carpenter, Kenward and Vansteelandt (2006), Davidian, Tsiatis and Leon (2005) and Bang and Robins (2005). What is not widely known, however, is that the methods developed by Robins et al. are not the only way to achieve double robustness. Many other types of estimates possess the DR property. Generalized regression estimators developed for sample surveys (Cassel, Särndal and Wretman, 1977), also known as model-assisted survey estimators (Särndal, Swensson and Wretman, 1989, 1992), have this property, as does a new class of parametric methods developed by Little and An (2004) and several other methods that do not seem to have been described before.

The first purpose of this article is pedagogical. We review a variety of incomplete-data estimation strategies, describe various ways in which the DR property can arise, and connect the recent articles on DR estimation to similar techniques found in the literature on sample surveys and causal inference. Our second purpose is to investigate the practical behavior of these estimators not only when either of the underlying models is correct, but in a scenario where both models are moderately misspecified.

### 1.2 Description of the Problem

For simplicity, we focus on the problem of estimating a population mean from an incomplete dataset. Many of the methods we present have been applied to the more general problem of estimating population-average regression coefficients, and we will describe those extensions where appropriate. Let us suppose that we have a random sample of units $i = 1, \ldots, n$ from an infinite population. The variable of primary interest is $y_i$. Let $t_i$ be the response indicator for $y_i$, so that $t_i = 1$ if $y_i$ is observed and $t_i = 0$ if $y_i$ is missing. For each unit, there is an observed $p$-dimensional vector of covariates $x_i$ that may be related both to $y_i$ and to $t_i$. A schematic representation of the sample data is shown in Figure 1(a). (In this figure, the indices $i = 1, \ldots, n$ have been permuted so that the sample units having $t_i = 1$ appear first.) Denote the numbers of respondents and nonrespondents by $n^{(1)} = \sum_i t_i$ and $n^{(0)} = \sum_i (1 - t_i)$, the population response and nonresponse rates by $r^{(1)} = P(t_i = 1)$ and $r^{(0)} = P(t_i = 0)$, and the sample rates by $\hat{r}^{(1)} = n^{(1)}/n$ and $\hat{r}^{(0)} = n^{(0)}/n$.

FIG. 1. *Schematic representation of sample data for estimating* (a) *a population mean and* (b) *an average causal effect, with missing values denoted by shading.*

The sample mean of the observed $y_i$'s,

$$\bar{y}^{(1)} = \frac{1}{n^{(1)}} \sum_i t_i y_i,$$

consistently estimates the mean for the respondents, $\mu^{(1)} = E(y_i \mid t_i = 1)$, under any reasonable population distribution and response mechanism. An estimator having this property will be said to be strongly robust. In general, there is no strongly robust estimate of the mean for the nonrespondents, $\mu^{(0)} = E(y_i \mid t_i = 0)$, or for the mean of the entire population, $\mu = E(y_i) = r^{(0)} \mu^{(0)} + r^{(1)} \mu^{(1)}$, based on the observed data alone. Naive estimates such as $\bar{y}^{(1)}$ may work well enough if $\hat{r}^{(0)}$ is small and the relationships among $x_i$, $t_i$ and $y_i$ are weak. As the nonresponse rate and the strength of these relationships grow, adjusting for selection bias becomes important, and inferences become sensitive to the assumptions underlying the adjustment.

This problem is closely related to estimating an average causal effect from an experiment or observational study. Suppose now that $t_i$ is an indicator of the treatment received by unit $i$. Associated with unit $i$ is a pair of potential outcomes: the response $y_{i1}$ that is realized if $t_i = 1$, and another response $y_{i0}$ that is realized if $t_i = 0$. This situation is depicted in Figure 1(b). The causal effect of the treatment on unit $i$, defined as $y_{i1} - y_{i0}$, is unobservable because one of the two potential outcomes is necessarily missing. It is often of interest to estimate the average causal effect (ACE) in the population,

$$ACE = E(y_{i1}) - E(y_{i0}),$$

or the ACE's among the treated and the untreated,

$$ACE^{(1)} = E(y_{i1} \mid t_i = 1) - E(y_{i0} \mid t_i = 1),$$
$$ACE^{(0)} = E(y_{i1} \mid t_i = 0) - E(y_{i0} \mid t_i = 0).$$

The notion of potential outcomes was introduced by Neyman (1923) for randomized experiments and by Rubin (1974a) for nonrandomized studies. Reviews of causal inference from this perspective are given by Holland (1986), Winship and Sobel (2004), Gelman and Meng (2004) and Rubin (2005). From Figure 1(b), it is apparent that the correlation between $y_{i1}$ and $y_{i0}$ [more precisely, the partial correlation between them given $x_i$; see Rubin (1974b)] cannot be estimated from the observed data. Without prior information on what this correlation might be—and it is unclear from where such information would come—one may separate the problem of estimating an ACE into independent estimation of the means of $y_{i1}$ and $y_{i0}$. Any of the methods described in this article can be used to estimate an average causal effect by applying the method separately to each potential outcome. When estimating ACE's, rates of missing information tend to be high and sensitivity to modeling assumptions may be acute.

### 1.3 Assumptions

Throughout this article, we will assume that the response mechanism is unconfounded in the sense that $y_i$ and $t_i$ are conditionally independent given $x_i$ (Rubin, 1978); this assumption has also been called strong ignorability (Rosenbaum and Rubin, 1983). Under strong ignorability, the joint distribution of the complete data can be written as

$$P(X,T,Y) = \prod_i P(x_i) P(t_i \mid x_i) P(y_i \mid x_i).$$

Strong ignorability implies that the missing $y_i$'s are missing at random (MAR) (Rubin, 1976; Little and Rubin, 2002). In many applications, MAR is unrealistic. Nevertheless, this assumption provides an important benchmark and point of departure for sensitivity analyses, and it is the foundation upon which DR procedures rest.

None of the methods we examine will require an explicit model for the covariates, but they will all make assumptions about $P(t_i \mid x_i)$, $P(y_i \mid x_i)$ or both. Denote the response probability for unit $i$ by

(1) $$P(t_i = 1 \mid x_i) = \pi_i(x_i) = \pi_i.$$

This probability is called the propensity score (Rosenbaum and Rubin, 1983). A proposed functional form for (1) will be called a $\pi$-model. If estimates of the propensity scores are needed, they are often taken to be

$$\hat{\pi}_i = \text{expit}(x_i^T \hat{\alpha}) = \frac{\exp(x_i^T \hat{\alpha})}{1 + \exp(x_i^T \hat{\alpha})},$$

where $\hat{\alpha}$ is the maximum-likelihood (ML) estimate of the coefficients from the logistic regression of $t_1, \ldots, t_n$ on $x_1, \ldots, x_n$. In many situations, of course, the assumed form of the $\pi$-model is not correct, and this misspecification can be problematic depending on how the $\hat{\pi}_i$'s are used.

Let us also define $E(y_i \mid x_i) = m(x_i) = m_i$, so that

(2) $$y_i = m(x_i) + \varepsilon_i$$

with $E(\varepsilon_i) = 0$. A functional form for $m(x_i)$ will be called a $y$-model. When an estimate of $m_i$ is needed, an obvious candidate is $\hat{m}_i = x_i^T \hat{\beta}$, where $\hat{\beta}$ is the vector of coefficients from the linear regression of $y_i$ on $x_i$ estimated from the respondents; $y$-models with nonlinear link functions are also straightforward (McCullagh and Nelder, 1989). In most cases, an analyst's regression model will be only a rough approximation to the true $y$-model. The implications of this misspecification can be serious if $P(x_i \mid t_i = 1)$ and $P(x_i \mid t_i = 0)$ are very different, because the $\hat{m}_i$'s for the nonrespondents will then be based on extrapolation. This is particularly true when $x_i$ is high-dimensional because of the so-called curse of dimensionality. With many covariates, it becomes difficult to specify a $y$-model that is sufficiently flexible to capture important nonlinear effects and interactions, yet parsimonious enough to keep the variance of prediction manageably low.

Under ignorability, the $\pi_i$'s play no role in likelihood-based or Bayesian analyses for the parameters of the $y$-model (Rubin, 1976). The parametric approach—specifying a full model for $y_i$ and ignoring the $\pi_i$'s—is emphasized in texts on missing data by Rubin (1987), Schafer (1997), Little and Rubin (2002) and others. Nevertheless, many advocates of the parametric approach also recognize that the $\pi_i$'s are useful for model validation and criticism (Gelman, Carlin, Rubin and Stern, 2004, Chapters 6–7). On the other hand, much of the literature on causal inference in the tradition of Rosenbaum and Rubin (1983) eschews models for $y_i$ in favor of matching and stratification based on propensity scores (e.g., Rosenbaum, 2002). The latter is motivated in part by a perceived inability to model the responses well enough to mitigate the dangers of extrapolation. Other uses for propensity scores, including inverse-propensity weighting (Robins, Rotnitzky and Zhao, 1994), also rely heavily on a $\pi$-model while relaxing assumptions about $y_i$. DR methods will require both a $y$-model and a $\pi$-model but remain consistent if one or the other is wrong. As we shall see, however, this property does not necessarily translate into improved performance when both models fail.

### 1.4 A Simulated Example

Consider the following example which, although artificial, bears some resemblance to what we have encountered in a real study. For each unit $i = 1, \ldots, n$, suppose that $(z_{i1}, z_{i2}, z_{i3}, z_{i4})^T$ is independently distributed as $N(0, I)$ where $I$ is the $4 \times 4$ identity matrix. The $y_i$'s are generated as

$$y_i = 210 + 27.4 z_{i1} + 13.7 z_{i2} + 13.7 z_{i3} + 13.7 z_{i4} + \varepsilon_i,$$

where $\varepsilon_i \sim N(0,1)$, and the true propensity scores are

$$\pi_i = \operatorname{expit}(-z_{i1} + 0.5 z_{i2} - 0.25 z_{i3} - 0.1 z_{i4}).$$

This mechanism produces an average response rate of $r^{(1)} = 0.5$, and the means are $\mu = 210.0$, $\mu^{(1)} = 200.0$ and $\mu^{(0)} = 220.0$. The selection bias in this example is not severe; the difference between the mean of the respondents and the mean of the full population is only one-quarter of a population standard deviation. Nevertheless, this difference is large enough to wreak havoc on the performance of the naive estimate $\bar{y}^{(1)}$ when used as the basis for confidence intervals and tests.

In this example, a logistic regression of $t_i$ on the $z_{ij}$'s would be a correct $\pi$-model, and a linear regression of $y_i$ on the $z_{ij}$'s would be a correct $y$-model. We will suppose, however, that instead of being given the $z_{ij}$'s, the covariates actually seen by the data analyst are

$$\begin{aligned}
x_{i1} &= \exp(z_{i1}/2), \\
x_{i2} &= z_{i2}/(1+\exp(z_{i1})) + 10, \\
x_{i3} &= (z_{i1} z_{i3}/25 + 0.6)^3, \\
x_{i4} &= (z_2 + z_4 + 20)^2.
\end{aligned}$$

This implies that $\operatorname{logit}(\pi_i)$ and $m_i$ are linear functions of $\log(x_{i1})$, $x_2$, $x_1^2 x_2$, $1/\log(x_1)$, $x_3/\log(x_1)$ and $x_4^{1/2}$. Except by divine revelation, it is unlikely that an analyst who sees only $x_i$ would ever formulate a correct $\pi$- or $y$-model. Rather, he or she would naturally be drawn to models that are linear and logistic in the $x_{ij}$'s, and those incorrect models look trustworthy. To illustrate, we drew a random sample of $n = 200$ units from this population, which happened to produce exactly $n^{(1)} = 100$ respondents and $n^{(0)} = 100$ nonrespondents. Scatterplots of $y_i$ versus $x_{ij}$, $j = 1, \ldots, 4$, for the 100 respondents are shown in Figure 2. Regressing $y_i$ on the $x_{ij}$'s yields coefficients for $x_{i1}$, $x_{i3}$ and $x_{i4}$ that are highly significant and a coefficient for $x_{i2}$ that is nearly significant at the 0.05-level; the prediction is strong ($R^2 = 0.81$), and a plot of residuals versus fitted values reveals no obvious outliers and little evidence of heteroscedasticity or nonlinearity. (In other samples, some higher-order terms such as $x_1^2$ and $x_1 x_2$, are significant and might be considered for inclusion in the model. Those terms, however, do little to improve the performance of any of the regression-based methods discussed below, and sometimes they are harmful.)

The covariates seen by the analyst are also related to the $t_i$'s. Side-by-side boxplots of the $x_{ij}$'s for the $t_i = 0$ and $t_i = 1$ groups are shown in Figure 3(a)–(d). Fitting a logistic model to this sample, we find that the coefficients for $x_{i1}$ and $x_{i2}$ are statistically significant, and all of the deviance residuals lie between $-1.85$ and $+2.51$. Figure 3(e) shows side-by-side boxplots of the linear predictors $\hat{\eta}_i = \operatorname{logit}(\hat{\pi}_i)$. As one would expect, the distributions of $\hat{\pi}_i$ in the two groups are different but not drastically so. To check the appropriateness of the link function, Hinkley (1985) suggested adding $\hat{\eta}_i^2$ as another covariate to see whether it is related to the response; the coefficient for this extra term was not significantly different from zero ($p = 0.20$), so in this sample an analyst would have little reason to alter the link.

Comparing the fitted values of $y_i$ under the true and misspecified $y$-models, we find that the correlation between them is approximately 0.9. Similarly, the correlation between the $\hat{\eta}_i$'s under the true and misspecified $\pi$-models is also about 0.9. This example appears to be precisely the type of situation for which the DR estimators of Robins et al. were developed. By relying on two reasonably good models, one hopes that at least one is close enough to the truth to yield satisfactory results. Indeed, Bang and Robins (2005, Section 2.1) state:

> In our opinion, a DR estimator has the following advantage that argues for its routine use: if either the [$y$-model] or the [$\pi$-model] is nearly correct, then the bias of a DR estimator of $\mu$ will be small. Thus, the DR estimator ... gives the analyst two chances to get nearly correct inferences about the mean of $Y$.

In Sections 2 and 3, we describe various techniques for estimating $\mu$ based on the observed data and evaluate their performance in this simulated example. Some of these methods use a $\pi$-model, some use a $y$-model, and some rely on both. All of the dual-modeling strategies possess a DR property, but they do not perform equally well. Pooling information from two models can be helpful, but the manner in which the information is pooled makes a difference. (Due to space limitations, we will not discuss computation of standard errors. Tractable variance estimates are available for most of these methods, but our purpose is to compare the performance of the estimates themselves.) In Section 4, we will provide further justification for why we constructed our example as we did, and we will outline the crucial differences between this and simulated examples used by Bang and Robins (2005) and others, so that apparently contradictory conclusions can be reconciled.

## 2. WEIGHTING, STRATIFICATION AND REGRESSION

### 2.1 Inverse-Propensity Weighting

Weighting observed values by inverse-probabilities of selection was proposed by Horvitz and Thompson (1952) in the context of survey inference for finite populations. The same idea is used in importance sampling, a Monte Carlo technique for approximating the moments of a distribution using random draws from another distribution that approximates it (Hammersley and Handscomb, 1964; Geweke, 1989). It is easy to see that $n^{-1}\sum_i t_i \pi_i^{-1} y_i$, which can be computed from the respondents alone, unbiasedly estimates the mean of the entire population, because strong ignorability implies that

$$E(t_i \pi_i^{-1} y_i) = E(E(t_i \pi_i^{-1} y_i \mid x_i))$$
$$= E(\pi_i \pi_i^{-1} m_i) = \mu.$$

Precision is often enhanced if we use a denominator of $\sum_i t_i \pi_i^{-1}$ rather than $n$, so that the estimate becomes a weighted average of the $y_i$'s for the respondents. Normalizing the weights in this manner, and replacing the unknown propensities by estimates derived from a $\pi$-model, the inverse-propensity weighted (IPW) estimate becomes

$$(3) \qquad \hat{\mu}_{IPW\text{-}POP} = \frac{\sum_i t_i \hat{\pi}_i^{-1} y_i}{\sum_i t_i \hat{\pi}_i^{-1}}.$$

The "$POP$" in the subscript indicates that we are reweighting the respondents to resemble the full population. Alternatively, we may reweight them to approximate the population of nonrespondents (Hirano and Imbens, 2001). An estimate of $\mu^{(0)}$ based on that idea is

$$\hat{\mu}_{IPW\text{-}NR}^{(0)} = \frac{\sum_i t_i \hat{\pi}_i^{-1}(1-\hat{\pi}) y_i}{\sum_i t_i \hat{\pi}_i^{-1}(1-\hat{\pi})},$$

and a corresponding estimate of $\mu$ is

$$(4) \qquad \hat{\mu}_{IPW\text{-}NR} = \hat{r}^{(1)} \bar{y}^{(1)} + \hat{r}^{(0)} \hat{\mu}_{IPW\text{-}NR}^{(0)}.$$

The IPW estimator can also be regarded as the solution to a simple weighted estimating equation $\sum_i w_i U_i = 0$, where $w_i = t_i \hat{\pi}_i^{-1}$ and $U_i = (y_i - \mu)/\sigma^2$ for any $\sigma^2 > 0$. From this standpoint, the IPW method can be generalized to estimate a vector of population-average coefficients for the regression of $y_i$ on an arbitrary set of covariates. In the regression setting, $U_i$ becomes a vector representing the $i$th unit's contribution to a quasi-score function. Coefficients estimated in this manner are consistent and asymptotically normally distributed provided that the $\pi_i$'s are bounded away from zero and the $\pi$-model has been correctly specified. Asymptotic properties and methods for variance estimation are described by Binder (1983) when the propensities are known, and by Robins, Rotnitzky and Zhao (1994, 1995) when the propensities have been estimated.

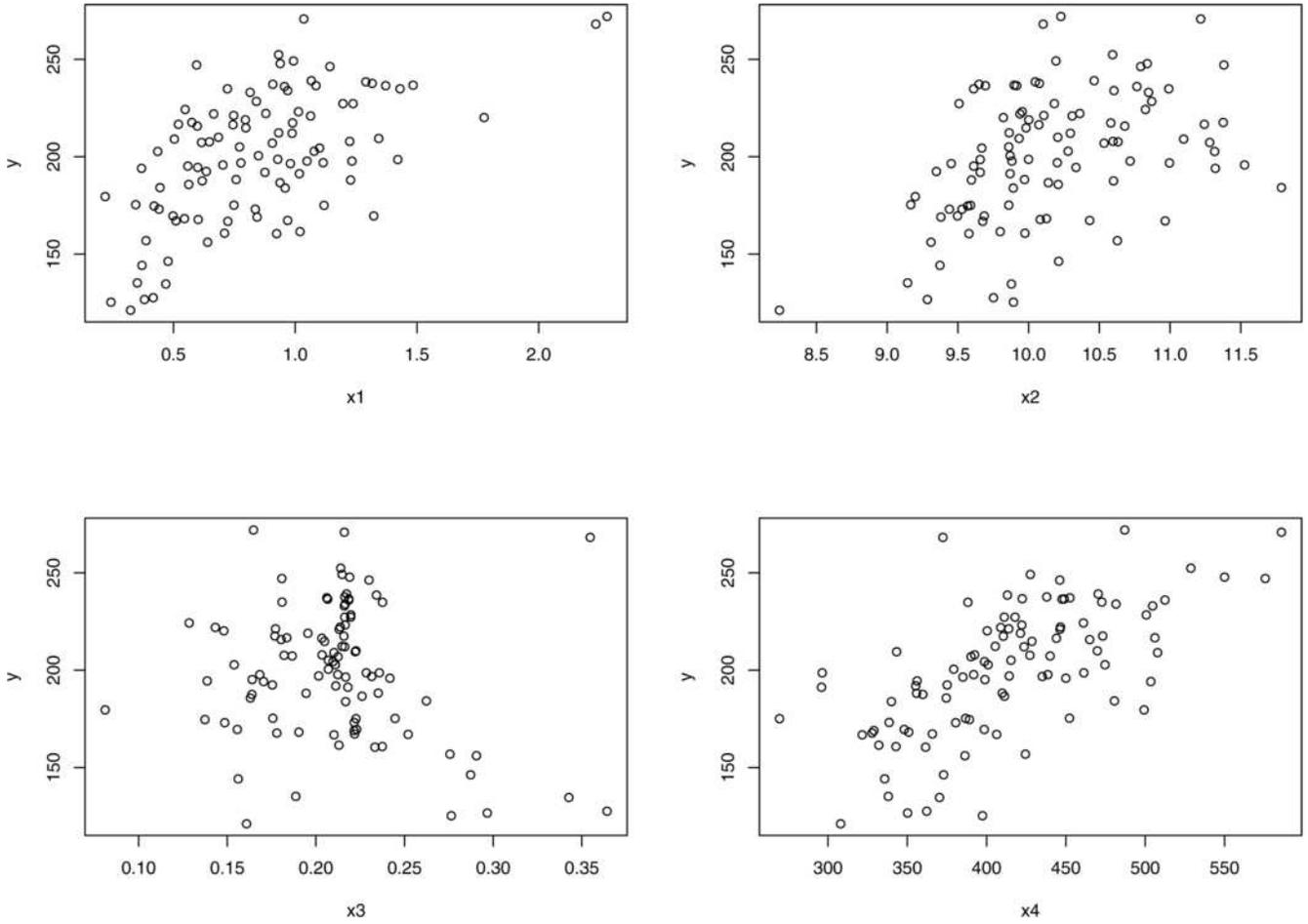

Fig. 2. *Scatterplots of response versus observed covariates for respondents in a sample of 200 units.*

In the original method of Horvitz and Thompson (1952), the $\pi_i$'s were determined by a known survey design. Surveys are usually designed to ensure that IPW estimates are acceptably precise, but the forces of nature that govern uncontrolled nonresponse are often unkind. In missing-data problems, IPW methods assign large weights to respondents who closely resemble nonrespondents, causing the estimates to have high variance. IPW estimates are also sensitive to misspecification of the $\pi$-model, because even mild lack of fit in outlying regions of the covariate space where $\pi_i \approx 0$ translates into large errors in the weights. These shortcomings of IPW estimators have been known for many years; see, for example, the comments of Little and Rubin (1987, Section 4.4.3) on the IPW methods for nonresponse proposed by Cassel, Särndal and Wretman (1983).

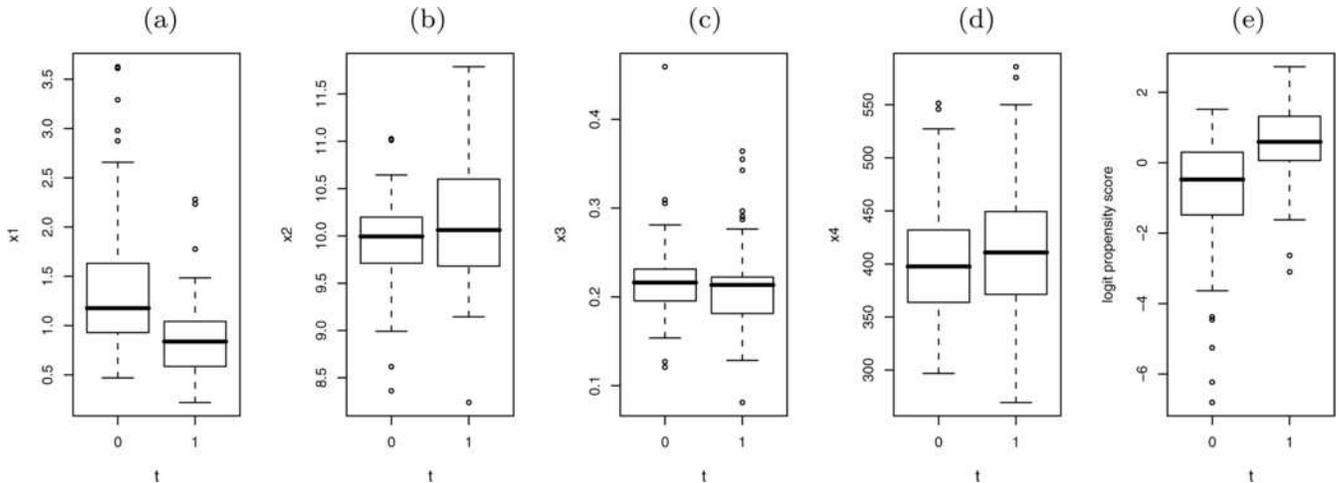

Fig. 3. *Distributions of* (a)–(d) *observed covariates and* (e) *estimated propensity scores for nonrespondents and respondents in a sample of 200 units.*

To see how well IPW performs on the artificial population described in Section 1.4, we created 1000 samples of $n = 200$ and $n = 1000$ units each and, for each sample, estimated the propensity scores in two ways. First, we fit a correctly specified $\pi$-model, regressing the $t_i$'s on $z_{i1}$, $z_{i2}$, $z_{i3}$ and $z_{i4}$ using a logit link. Second, we fit the incorrect model which replaces the $z_{ij}$'s with $x_{ij}$'s. The behavior of the four IPW estimates (3)–(4) under the correctly and incorrectly specified $\pi$-models is summarized in Table 1. In this table, "Bias" is the average difference between the estimate and $\mu = 210$, and "% Bias" is the bias as a percentage of the estimate's standard deviation. (A useful rule of thumb is that the performance of interval estimates and test statistics begins to deteriorate when the bias of the point estimate exceeds about 40% of its standard deviation.) "RMSE" is square root of the average value of $(\hat{\mu} - \mu)^2$. Examining Table 1, we see that the IPW estimates are biased when the $\pi$-model is misspecified, and the biases are accompanied by huge losses in precision. In fact, the bias and RMSE actually get worse as the sample size grows! IPW estimates have higher variance than other procedures examined in this article even when the propensities are correctly modeled, but when the $\pi$-model is incorrect, the method breaks down. Interestingly, NR weighting performs better than POP weighting; we do not know whether the superiority of NR is a peculiar feature of this example or if it tends to hold more generally.

The poor performance of IPW is due in part to occasional highly erratic estimates produced by a few enormous weights. In practice, a good data analyst would never use a simple IPW estimator if the weights were too extreme. Unusually large weights may be taken as a sign of model failure, prompting the researcher to revise the $\pi$-model. Outlying weights may be truncated or shrunk to more sensible values, or the offending units with large weights may be removed. (Removing these units is not recommended, because the respondents with the lowest propensities are in fact those that contain the best information for predicting nonrespondent behavior.) Analysts who apply IPW to real problems quickly learn that it often cannot be used without ad hoc modifications. The column of Table 1 labeled "MAE" reports the median of the absolute errors $|\hat{\mu} - \mu|$, which discards the worst 50% of the estimates. Even by this robust measure of precision, IPW performs more poorly than the other methods we examine when the $\pi$-model is misspecified, and it fails to improve as the sample size grows.

In many applications of IPW methodology, weights are obtained by logistic regression. Logistic models can be a poor way to estimate response propensities, because ML estimates from the logistic model are not resistant to outliers (Pregibon, 1982). A promising alternative is the robit model, which replaces the logistic link by the cumulative distribution function of a Student's t-distribution with $\nu$ degrees of freedom (Albert and Chib, 1993). The logit link is well approximated by the robit with $\nu = 7$, and smaller values of $\nu$ lead to estimates that are more robust (Liu, 2004). In this example, robit models produce minor improvements when the covariates are correct and major improvement when the covariates are wrong. We found that, with samples of $n = 1000$, replacing the logit link by robit with $\nu = 4$ reduces the bias and RMSE by nearly 50% when the $\pi$-model is incorrect. If IPW must be used, replacing the logistic regression with a more robust procedure can be advantageous.

TABLE 1
*Performance of IPW estimators of $\mu$ over 1000 samples from the artificial population*

| Sample size | $\pi$-model | Method | Bias | % Bias | RMSE | MAE |
|---|---|---|---|---|---|---|
| (a) $n = 200$ | Correct | *IPW-POP* | −0.27 | −7.0 | 3.86 | 2.43 |
| | | *IPW-NR* | −0.29 | −8.2 | 3.60 | 2.36 |
| | Incorrect | *IPW-POP* | 1.58 | 19.2 | 8.35 | 3.32 |
| | | *IPW-NR* | 0.61 | 10.3 | 5.99 | 3.03 |
| (b) $n = 1000$ | Correct | *IPW-POP* | −0.01 | −0.5 | 1.81 | 1.16 |
| | | *IPW-NR* | −0.03 | −1.5 | 1.68 | 1.09 |
| | Incorrect | *IPW-POP* | 5.05 | 45.9 | 12.10 | 2.80 |
| | | *IPW-NR* | 3.22 | 49.1 | 7.29 | 2.34 |

### 2.2 Propensity-Based Stratification

To mitigate the dangers of extreme weights and misspecification of the $\pi$-model, some prefer to coarsen the estimated propensity scores into a few categories and compute weighted averages of the mean response across categories. In the context of survey nonresponse, the technique is known as weighting-cell estimation or adjustment (Oh and Scheuren, 1983; Little, 1986; Little and Rubin, 2002). Strong ignorability implies that $y_i$ and $t_i$ are conditionally independent within any subpopulation for which $\pi_i(x_i)$ is constant (Rosenbaum and Rubin, 1983). In classes of constant propensity, the mean values of $y_i$ for respondents and nonrespondents are equal, which implies that

$$(5) \qquad \mu = \int E(y_i \mid \pi_i, t_i = 1) \, dP(\pi_i).$$

Suppose we fit a $\pi$-model and define strata $s = 1, \ldots, S$ by grouping units whose $\hat{\pi}_i$'s are similar. Define $c_{is} = 1$ if unit $i$ belongs to stratum $s$ and 0 otherwise. The $\hat{\pi}$-stratified estimate of $\mu$ approximates (5) by a weighted average of respondents' mean in each stratum, weighted by the proportion of sample units in that stratum,

$$(6) \qquad \hat{\mu}_{strat\text{-}\pi} = \sum_{s=1}^{S} \left( \frac{\sum_i c_{is}}{n} \right) \left( \frac{\sum_i c_{is} t_i y_i}{\sum_i c_{is} t_i} \right).$$

Similarly, a $\hat{\pi}$-stratified estimate of $\mu^{(0)}$ weights the respondents' mean in each stratum by the proportion of nonrespondents in that stratum,

$$\hat{\mu}_{strat\text{-}\pi}^{(0)} = \sum_{s=1}^{S} \left( \frac{\sum_i c_{is}(1-t_i)}{n^{(0)}} \right) \left( \frac{\sum_i c_{is} t_i y_i}{\sum_i c_{is} t_i} \right),$$

which may be combined with $\bar{y}^{(1)}$ as in (4) to produce another estimate of $\hat{\mu}$. Rosenbaum and Rubin (1983) suggest classifying units into $S = 5$ strata defined by the sample quintiles of $\hat{\pi}_i$, as this tends to eliminate more than 90% of the selection bias if the $\pi$-model is correct (Cochran, 1968).

The performance of the $\hat{\pi}$-stratified estimator (6) over the 1000 samples from the artificial population is summarized in Table 2. Comparing these results to those of Table 1, we see that stratification is less effective than IPW at removing bias when the $\pi$-model is correct, but the stratified estimators still outperform IPW in terms of RMSE. The increase in bias incurred by coarsening the $\hat{\pi}_i$'s is easily offset by the greater efficiency that results from stabilizing the largest weights. When the $\pi$-model is misspecified, the stratified estimators are more biased than IPW for samples of $n = 200$ but less biased when $n = 1000$, because the bias of the stratified estimators does not worsen as $n$ increases.

### 2.3 Regression Estimation

IPW and $\hat{\pi}$-stratified estimators pay no heed to relationships between the covariates and $y_i$. Regression estimators, on the other hand, model $y_i$ from $x_i$ directly and use this information to predict the missing values. Because strong ignorability implies that

$$E(y_i \mid x_i, t_i = 0) = E(y_i \mid x_i, t_i = 1) = E(y_i \mid x_i),$$

we can regress $y_i$ on $x_i$ among the respondents, apply the estimated regression function to predict $y_i$ for the entire sample, and then average the predicted values to obtain an estimate of $\mu$. Let

$$\hat{\beta} = \left( \sum_j t_j x_j x_j^T \right)^{-1} \left( \sum_j t_j x_j y_j \right)$$

denote the ordinary least-squares (OLS) coefficients from the regression of $y_i$ on $x_i$ among the respondents, and let $\hat{m}_i = x_i^T \hat{\beta}$. The OLS regression estimate for $\mu$ is

$$(7) \qquad \hat{\mu}_{OLS} = \frac{1}{n} \sum_i \hat{m}_i.$$

This estimate is unbiased if the $y$-model is true, that is, if $E(y_i \mid x_i) = x_i^T \beta$ for some $\beta \in \mathcal{R}^p$; in addition, it is highly efficient if $\sigma_i^2 = V(y_i \mid x_i)$ is nearly constant. If the response is heteroscedastic, efficiency can be improved by replacing $\hat{\beta}$ with a weighted least-squares estimate with weights proportional to $\sigma_i^{-2}$.

From our 1000 samples, we computed OLS regression estimates under a correct $y$-model (regressing $y_i$ on the $z_{ij}$'s) and an incorrect $y$-model (regressing $y_i$ on the $x_{ij}$'s). The results, which are summarized in Table 3, verify that the bias is indeed removed when the $y$-model is correct but not when the model is wrong. Comparing the RMSE values in this table with those in Tables 1 and 2, we see that estimates based on the incorrect $y$-model are more stable and efficient than those based on the incorrect $\pi$-model. The bias that remains due to misspecification of the $y$-model is not large in absolute terms, but it is still troubling in samples of $n = 1000$ because there it amounts to more than 50% of a standard error and begins to impair tests and intervals. Difficulties with parametric missing-data methods arise when the uncertainty due to the model specification, which is rarely accounted for, grows relative to the sampling variation under the assumed model. In those cases, a point estimate based on a misspecified but reasonable $y$-model may still perform better than other estimates, but the analyst is too optimistic about its precision.

The performance of the regression estimate depends heavily on the strength of the correlation $R$ between $y_i$ and $\hat{m}_i$. As $R^2$ approaches 0, $\hat{\mu}_{OLS}$ converges to $\bar{y}^{(1)}$ and suffers from the same bias as this naive estimate. As $R^2$ approaches 1, it converges to the mean of the full sample, $\bar{y} = n^{-1} \sum_i y_i$, which is strongly robust. Therefore, if the $y$-model has strong prediction, the regression estimator tends to dominate other methods in terms of bias, efficiency and robustness. The IPW and $\pi$-stratified estimators, on the other hand, break down as the predictive ability of the $\pi$-model increases.

### 2.4 Stratification by Propensity Scores and Predicted Values

We saw that the efficiency and robustness of an IPW estimator can be enhanced by coarsening the $\hat{\pi}_i$'s into a small number of categories. The efficiency of these estimators can be further increased by adding covariates to the $\pi$-model that are predictive of $y_i$, even if these covariates are unrelated to $t_i$ (Lunceford and Davidian, 2004). Vartivarian and Little (2002) show that further improvement is possible if we cross-classify units by estimated propensity scores and covariates that are strongly related to the outcome. The idea of combining estimated propensity scores with additional covariate information is not new. For example, Rosenbaum and Rubin (1985) recommended matching respondents to nonrespondents or vice versa by a Mahalanobis-distance criterion based on $x_i$ within calipers defined by the estimated propensity scores. Numerous authors have performed regression adjustments based on $x_i$ within strata defined by $\hat{\pi}$; for references, see D'Agostino (1998).

To illustrate a simple version of this idea, suppose that we create 25 strata by cross-classifying units into cells defined by the sample quintiles of $\hat{\pi}_i$ and $\hat{m}_i$, and then estimate $\mu$ by a weighted average of the mean response across the cells. If we apply this procedure over repeated samples, we will occasionally encounter a cell that contains nonrespondents but no respondents, in which case the stratified estimates are undefined. For those samples, we must modify the estimate in some fashion. For example, we may collapse adjacent rows or columns in the $5 \times 5$ table, fuse the offending cell with an adjacent cell, impute the missing cell mean by a regression estimate that assumes the response surface has row and column effects but no interactions, and so on.

TABLE 2
*Performance of propensity-stratified estimators over 1000 samples from the artificial population*

| Sample size | $\pi$-model | Method | Bias | % Bias | RMSE | MAE |
|---|---|---|---|---|---|---|
| (a) $n = 200$ | Correct | $strat$-$\pi$ | −1.15 | −38.1 | 3.22 | 2.17 |
| | Incorrect | $strat$-$\pi$ | −2.82 | −87.7 | 4.28 | 3.13 |
| (b) $n = 1000$ | Correct | $strat$-$\pi$ | −1.08 | −81.5 | 1.71 | 1.18 |
| | Incorrect | $strat$-$\pi$ | −2.87 | −202.7 | 3.19 | 2.83 |

TABLE 3
*Performance of ordinary least-squares regression estimators over 1000 samples from the artificial population*

| Sample size | $y$-model | Method | Bias | % Bias | RMSE | MAE |
|---|---|---|---|---|---|---|
| (a) $n = 200$ | Correct | $OLS$ | −0.08 | −3.4 | 2.48 | 1.68 |
| | Incorrect | $OLS$ | −0.57 | −17.7 | 3.26 | 2.24 |
| (b) $n = 1000$ | Correct | $OLS$ | −0.00 | −0.1 | 1.17 | 0.79 |
| | Incorrect | $OLS$ | −0.84 | −56.0 | 1.72 | 1.15 |

In general, we dislike procedures that require frequent ad hoc adjustments unless their operating characteristics are well understood and the variance of the adaptive estimator can be approximated. Nevertheless, it is interesting to see how well the method performs in our artificial example. For each of our samples, we computed $(\hat{\pi} \times \hat{m})$-stratified estimators using all four possible combinations of correct and incorrect $\pi$- and $y$-models. Missing cell means were imputed by a regression procedure that assumes no row-by-column interactions. The results, which are summarized in Table 4, show that dual stratification produces a crude kind of double robustness; the bias is relatively low if the $\pi$-model is correctly specified or if the $y$-model is correctly specified. Under strong ignorability, a stratified estimate of the population mean will be unbiased if either the true $\pi_i$'s or the true $m_i$'s are constant within strata.

It is also useful to compare the results in Table 4 where both models are incorrect with those of Table 2 where the $\pi$-model is incorrect. This comparison shows that a $\pi$-stratified estimator can be improved with predicted values for the $y_i$'s, even if those predictions come from a coarsened, misspecified $y$-model. The key idea of DR estimation—reducing your reliance on one model by specifying two—does produce modest gains in this example over estimates based on a $\pi$-model alone. Comparing Tables 3 and 4, however, we find that the approximate DR procedure based on two incorrect models performs worse than OLS based on the incorrect $y$-model. When neither model is exactly true, two models are not necessarily better than one.

## 3. CONSTRUCTING DOUBLY ROBUST ESTIMATES

### 3.1 Regression Estimation with Residual Bias Correction

Cassel, Särndal and Wretman (1976, 1977) introduced a family of "generalized regression estimators" for population means and totals that combine model-based predictions for $y_i$ with inverse-probability weights. These methods, which are part of a methodology called model-assisted survey estimation (Särndal, Swensson and Wretman, 1992), are highly efficient when the $y$-model is true, yet remain asymptotically unbiased when the $y$-model is misspecified. In the original formulation, the response probabilities were a known feature of the survey design, but with uncontrolled nonresponse the propensities may be estimated under a $\pi$-model (Cassel, Särndal and Wretman, 1983).

To understand how these generalized regression estimators work, consider the simple regression estimator (7). If the regression model holds in the sense that $E(y_i|x_i) = x_i^T \beta$ for some $\beta \in \mathcal{R}^p$, then on average the predictions $\hat{m}_i = x_i^T \hat{\beta}$ will be neither too high nor too low; the mean of the estimated residuals $\hat{\varepsilon}_i = y_i - \hat{m}_i$ in the population will be zero. Of course, residuals are seen only for sampled respondents. We can, however, consistently estimate the mean residual for the full population if we have access to a $\pi$-model, and this estimate can in turn be used to correct the OLS estimate for bias arising from $y$-model failure. Cassel, Särndal and Wretman (1976) proposed the bias-corrected estimate

$$(8) \qquad \hat{\mu}_{BC\text{-}OLS} = \hat{\mu}_{OLS} + \frac{1}{n} \sum_i t_i \hat{\pi}_i^{-1} \hat{\varepsilon}_i.$$

Notice that if the $y$-model is true, then $E(\hat{\varepsilon}_i) = 0$, and the second term on the right-hand side of (8) has expectation zero for arbitrary $\hat{\pi}_i$'s. If the $\pi$-model is true, then the second term consistently estimates (minus one times) the bias of the first term. Therefore, this estimate is doubly robust.

Many variations on this approach are possible. For example, we can normalize the weights in the correction term, so that the estimate becomes

$$\hat{\mu}_{OLS} + \frac{\sum_i t_i \hat{\pi}^{-1} \hat{\varepsilon}_i}{\sum_i t_i \hat{\pi}^{-1}}.$$

Or we can replace the POP weights with NR weights, so that the correction term estimates the mean resid-

TABLE 4
*Performance of propensity and fitted-value stratified estimators over 1000 samples from the artificial population*

| Sample size | $\pi$-model | $y$-model | Method | Bias | % Bias | RMSE | MAE |
|---|---|---|---|---|---|---|---|
| (a) $n = 200$ | Correct | Correct | strat-$\pi m$ | $-0.34$ | $-11.6$ | 2.92 | 1.90 |
| | | Incorrect | strat-$\pi m$ | $-0.59$ | $-18.4$ | 3.25 | 2.19 |
| | Incorrect | Correct | strat-$\pi m$ | $-0.49$ | $-17.1$ | 2.89 | 1.96 |
| | | Incorrect | strat-$\pi m$ | $-2.00$ | $-61.4$ | 3.82 | 2.62 |
| (b) $n = 1000$ | Correct | Correct | strat-$\pi m$ | $-0.27$ | $-21.1$ | 1.31 | 0.87 |
| | | Incorrect | strat-$\pi m$ | $-0.45$ | $-33.7$ | 1.42 | 0.92 |
| | Incorrect | Correct | strat-$\pi m$ | $-0.51$ | $-39.6$ | 1.38 | 0.92 |
| | | Incorrect | strat-$\pi m$ | $-2.10$ | $-148.7$ | 2.53 | 2.11 |

ual in the population of nonrespondents. A bias-corrected estimate of $\mu^{(0)}$ based on this idea is

$$\frac{\sum_i (1-t_i) x_i^T \hat{\beta}}{\sum_i (1-t_i)} + \frac{\sum_i t_i \hat{\pi}^{-1}(1-\hat{\pi}_i) \hat{\varepsilon}_i}{\sum_i t_i \hat{\pi}^{-1}(1-\hat{\pi}_i)},$$

which can be combined with $\bar{y}^{(1)}$ as in (4) to produce another DR estimate for $\hat{\mu}$. A third possibility is to replace the weighted correction term with a $\hat{\pi}$-stratified estimate of $E(\hat{\varepsilon}_i)$. Using five strata based on sample quintiles of $\hat{\pi}_i$ would remove over 90% of the bias from the OLS estimate if the $\pi$-model were true and reduce problems of instability caused by very large weights.

A more general version of (8) was independently proposed by Robins, Rotnitzky and Zhao (1994) for estimating population-average regression coefficients from incomplete data. Suppose $U_i$ is the contribution of sample unit $i$ to a vector-valued quasi-score function for the regression of $y_i$ on an arbitrary set of covariates. As noted in Section 2.1, the solution to $\sum_i w_i U_i = 0$ with $w_i = t_i \hat{\pi}_i^{-1}$ provides a consistent and asymptotically normal estimate of the population-average regression coefficients if the model used to estimate the $\pi_i$'s is correct. Now suppose that we change the estimating equations to $\sum_i [w_i U_i + (1-w_i) \phi_i] = 0$, where $\phi_i = \phi(x_i)$ is an arbitrary term that may depend on $x_i$ but not on $y_i$. The mean of the additional term $(1-w_i)\phi_i$ is essentially zero if the $\pi$-model is true, because $E(w_i) = E(E(w_i \mid x_i)) = E(\hat{\pi}_i^{-1} \pi_i) \approx 1$. Therefore, the solution to these augmented inverse-probability weighted (AIPW) estimating equations is again consistent and asymptotically normally distributed under a correct $\pi$-model. Robins and Rotnitzky (1995) demonstrate that a judicious choice for $\phi_i$ can greatly improve upon the efficiency of the simple IPW estimator. In particular, choosing $\phi_i = E(U_i \mid x_i)$, where the expectation is taken with respect to the distribution for $y_i$ given $x_i$, produces a locally semiparametric efficient estimator, the most efficient estimator within this class. This estimate is DR, maintaining its consistency if either the $\pi$-model or $y$-model is correct (Scharfstein, Rotnitzky and Robins, 1999). Taking $U_i = (y_i - \mu)/\sigma^2$, and estimating $E(y_i \mid x_i)$ by $(\hat{m}_i - \mu)/\sigma^2$ where $\hat{m}_i = x_i^T \hat{\beta}$, the solution to the AIPW estimating equation reduces to the generalized regression estimator (8). More generally, it becomes the solution to

$$(9) \qquad \frac{1}{n}\sum_i \hat{U}_i + \frac{1}{n}\sum_i t_i \hat{\pi}_i^{-1}(U_i - \hat{U}_i) = 0,$$

where $\hat{U}_i$ is the quasi-score function $U_i$ with $y_i$ replaced by $\hat{m}_i$.

TABLE 5
*Performance of bias-corrected regression estimators over 1000 samples from the artificial population*

| Sample size | $\pi$-model | $y$-model | Method | Bias | % Bias | RMSE | MAE |
|---|---|---|---|---|---|---|---|
| (a) $n = 200$ | Correct | Correct | BC-OLS | $-0.08$ | $-3.4$ | 2.48 | 1.68 |
| | | Incorrect | BC-OLS | 0.25 | 7.5 | 3.28 | 2.17 |
| | Incorrect | Correct | BC-OLS | $-0.08$ | $-3.3$ | 2.48 | 1.70 |
| | | Incorrect | BC-OLS | $-5.12$ | $-43.0$ | 12.96 | 3.54 |
| (b) $n = 1000$ | Correct | Correct | BC-OLS | 0.00 | $-0.1$ | 1.17 | 0.79 |
| | | Incorrect | BC-OLS | 0.06 | 3.4 | 1.75 | 1.02 |
| | Incorrect | Correct | BC-OLS | $-0.02$ | $-1.4$ | 1.49 | 0.80 |
| | | Incorrect | BC-OLS | $-21.03$ | $-13.5$ | 157.21 | 5.32 |

By analogy to (8), (9) can be viewed as a predicted estimating equation with a residual bias correction. Any of the variations on (8) described above—normalizing the weights, switching to NR weights, or switching from an IPW bias correction to a $\hat{\pi}$-stratified one—can be applied to (9) as well. Modifications like these would take the estimator outside of the AIPW class. Nevertheless, these changes could potentially improve performance when either or both models are misspecified.

The performance of the bias-corrected regression estimate (8) in our simulated example is summarized in Table 5. The bias of this estimate does indeed vanish when either of the two models is correct. Moreover, comparing these results to those from the IPW-POP estimates in Table 1, we see that augmenting the IPW procedure by information from a correct $y$-model does indeed increase the efficiency. In the more realistic condition where both models are misspecified, however, this DR estimate does worse than IPW. A similar pattern emerges if we compare the new results to those from the simple OLS regression estimate in Table 3. Bias from an incorrect $y$-model is repaired by a $\pi$-model if the latter is correct. When both models are misspecified, however, the DR procedure is substantially worse than OLS. Like IPW, estimates from this DR procedure often behave erratically because one or more weights are occasionally enormous. Even if we trim away these "bad" samples and judge the performance by the MAE, however, the new procedure is still worse than IPW and OLS. Once again, two models are not necessarily better than one.

Why does this DR estimate fail to perform better than IPW and OLS even though the $\pi$- and $y$-models are reasonably close to being true? The local semiparametric efficiency property, which guarantees that the solution to (9) is the best estimator within its class, was derived under the assumption that both models are correct. This estimate is indeed highly efficient when the $\pi$-model is true and the $y$-model is highly predictive. In our experience, however, if the $\pi$-model is misspecified, it is not difficult to find DR estimators that outperform this one by venturing outside the AIPW class. For this particular example, normalizing the POP weights, switching to NR weights, and using a $\hat{\pi}$-stratified bias correction all improve upon $\hat{\mu}_{BC\text{-}OLS}$. There are yet more ways to construct DR estimates, as we now describe.

### 3.2 Regression Estimation with Inverse-Propensity Weighted Coefficients

The correction term in $\hat{\mu}_{BC\text{-}OLS}$ repairs the bias in $\hat{\mu}_{OLS} = n^{-1}\sum_i x_i^T \hat{\beta}$ by estimating the mean residual in the full population. A different way to repair this bias is to move the estimated coefficients away from $\hat{\beta}$. Imagine that we could see the OLS coefficients based on the full sample,

$$\hat{\beta}_{SAMP} = \left(\sum_i x_i x_i^T\right)^{-1}\left(\sum_i x_i y_i\right).$$

A well-known property of OLS regression is that the sum of the estimated residuals $y_i - x_i^T \hat{\beta}_{SAMP}$, $i = 1,\ldots,n$, is zero if $x_i$ includes a constant. This is an algebraic identity that holds regardless of the actual form of $E(y_i \mid x_i)$. This identity implies that the regression estimator based on $\hat{\beta}_{SAMP}$ replicates the mean of $y_i$ in the full sample,

$$\frac{1}{n}\sum_i x_i^T \hat{\beta}_{SAMP} = \frac{1}{n}\sum_i y_i = \bar{y},$$

which is a strongly robust estimate of $\mu$. We cannot compute $\hat{\beta}_{SAMP}$ from the observed data. But with propensities estimated from a $\pi$-model, we can compute a weighted least-squares (WLS) estimate

$$\hat{\beta}_{WLS} = \left(\sum_i t_i \hat{\pi}_i^{-1} x_i x_i^T\right)^{-1}\left(\sum_i t_i \hat{\pi}_i^{-1} x_i y_i\right).$$

In a well-behaved asymptotic sequence, $\hat{\beta}_{SAMP}$ and $\hat{\beta}_{WLS}$ both converge to the coefficients from the linear regression of $y_i$ on $x_i$ in the full population, regardless of whether that regression is an accurate portrayal of $E(y_i \mid x_i)$. If we compute a regression estimate for $\mu$ based on the WLS coefficients,

$$(10) \qquad \hat{\mu}_{WLS} = \frac{1}{n}\sum_i x_i^T \hat{\beta}_{WLS},$$

the difference between this estimate and $\bar{y}$ converges in probability to zero as $n \to \infty$, provided that the $\pi$-model holds.

From this discussion, we see that $\hat{\mu}_{WLS}$ consistently estimates $\mu$ if the $\pi$-model is true. If the $y$-model is true, then $\hat{\beta}_{WLS}$ will be an inefficient but consistent estimate of $\beta$, and $\hat{\mu}_{WLS}$ will again be consistent; thus it is DR.

In our simulated example, the WLS regression estimate is sometimes inferior to the bias-corrected OLS estimate (8) when one of the models is true, but much better when both models are misspecified (Table 6). Comparing $\hat{\mu}_{WLS}$ to $\hat{\mu}_{OLS}$, we see that the inverse-propensity weighted estimate of $\beta$ effectively corrects the bias from a wrong $y$-model if the $\pi$-model is correct, but makes matters worse if the $\pi$-model is wrong.

Once again, many variations on (10) are possible. Normalizing the POP weights $t_i \hat{\pi}_i^{-1}$ has no effect on $\hat{\beta}_{WLS}$, but one might consider switching to NR weights. Applying NR weights and averaging the predicted values of $x_i^T \beta$ among nonrespondents produces an estimate of $\hat{\mu}^{(0)}$, which can then be combined with $\bar{y}^{(1)}$ to produce another estimate of $\mu$. Another possibility is to coarsen the weights into, say, five categories and compute a $\hat{\pi}$-stratified estimate of $\beta$.

### 3.3 Regression Estimation with Propensity-Based Covariates

A third general strategy for constructing a DR estimate is to incorporate functions of estimated propensities into the $y$-model as covariates. Ordinary regression estimates are based on the relationship

$$\mu = \int E(y_i \mid x_i) \, dP(x_i)$$

and achieve consistency if $E(y_i \mid x_i) = E(y_i \mid x_i, t_i = 1)$ is consistently estimated for all $x_i$. In prac-tice, creating a model that gives unbiased predictions for $y_i$ over the whole covariate space can be a daunting task, because real data often exhibit non-linearities, interactions, etc. that are difficult to identify and portray, especially as the dimension of $x_i$ grows. From (5), however, we see that requiring unbiased prediction for all $x_i$ is much stronger than necessary; it would suffice to have unbiased prediction of $E(y_i \mid \pi(x_i)) = E(y_i \mid \pi(x_i), t_i = 1)$ for all $\pi(x_i) \in (0, 1)$. If we want to repair the bias in a parametric $y$-model, it makes sense to first identify and correct for lack of fit in the direction of the propensity score, because $\pi_i = \pi(x_i)$ is the coarsest summary of $x_i$ that makes $y_i$ and $t_i$ conditionally independent (Rosenbaum and Rubin, 1983).

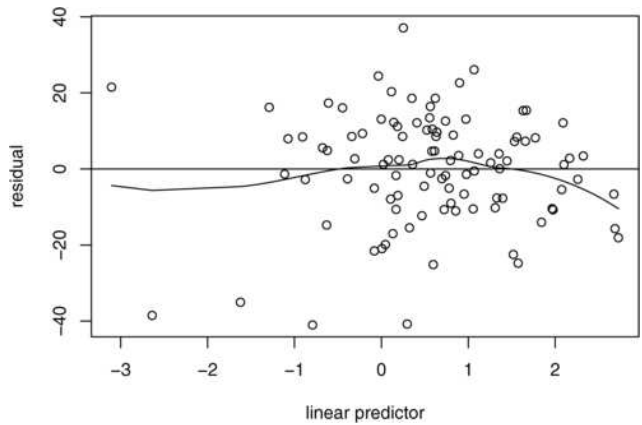

FIG. 4. *Scatterplot of raw residuals from $y$-model against linear predictors from a $\pi$-model, with least-squares and local polynomial fit.*

Consider the sample of $n = 200$ units from our artificial population that we examined in Section 1.4. Figure 4 shows the residuals $\hat{\varepsilon}_i$ from the linear regression of $y_i$ on $x_i$, plotted against the linear predictors $\hat{\eta}_i$ from the logistic regression of $t_i$ on $x_i$, for the $n^{(1)} = 100$ responding units. A least-squares line fit to this plot has an intercept and slope of zero, because the predictor is a perfect linear combination of covariates already in the $y$-model. A smooth curve created by a local polynomial (loess) fit, however, suggests that predictions from the $y$-model tend to be slightly low in the middle of the propensity scale and slightly high at the extremes. The bias could be corrected by fitting a generalized additive model (Hastie and Tibshirani, 1990) that allows the mean of $y_i$ to vary smoothly with $\hat{\pi}_i$ in a nonparametric fashion. Little and An (2004) incorporated a smoothing spline based on $\hat{\eta}_i$ and demonstrated that the resulting regression estimate of $\mu$ is DR in the following sense: If the $y$-model correctly describes $E(y_i \mid x_i)$ before the propensity-related terms are added, these additional terms (or any other functions of $x_i$) merely cause the model to be overfitted and add mean-zero noise to the predicted values. If the $\pi$-model is correct, then (5) guarantees consistent estimation of $\mu$, as long as the mean of $y_i$ varies smoothly with $\pi_i$ and this relationship can be arbitrarily well approximated by the linear combination of basis functions added to the model. In the latter case, the propensity-related covariates completely remove the bias for estimating $\mu$, and any additional information provided by $x_i$ merely serves to make the estimate more precise.

TABLE 6
*Performance of regression estimators with inverse-propensity weighted coefficients over 1000 samples from the artificial population*

| Sample size | $\pi$-model | $y$-model | Method | Bias | % Bias | RMSE | MAE |
|---|---|---|---|---|---|---|---|
| (a) $n = 200$ | Correct | Correct | WLS | −0.09 | −3.4 | 2.48 | 1.68 |
| | | Incorrect | WLS | 0.38 | 13.2 | 2.88 | 1.92 |
| | Incorrect | Correct | WLS | −0.08 | −3.4 | 2.48 | 1.68 |
| | | Incorrect | WLS | −2.20 | −70.0 | 3.83 | 2.74 |
| (b) $n = 1000$ | Correct | Correct | WLS | 0.00 | −0.1 | 1.17 | 0.78 |
| | | Incorrect | WLS | 0.16 | 12.0 | 1.35 | 0.92 |
| | Incorrect | Correct | WLS | 0.00 | −0.1 | 1.17 | 0.78 |
| | | Incorrect | WLS | −2.99 | −203.6 | 3.33 | 2.98 |

TABLE 7
*Performance of propensity-covariate (four dummy indicators) regression estimators over 1000 samples from the artificial population*

| Sample size | $\pi$-model | $y$-model | Method | Bias | % Bias | RMSE | MAE |
|---|---|---|---|---|---|---|---|
| (a) $n = 200$ | Correct | Correct | $\pi$-*cov* | −0.09 | −3.4 | 2.48 | 1.69 |
| | | Incorrect | $\pi$-*cov* | −0.39 | −13.5 | 2.93 | 2.00 |
| | Incorrect | Correct | $\pi$-*cov* | −0.09 | −3.4 | 2.48 | 1.68 |
| | | Incorrect | $\pi$-*cov* | −1.27 | −38.6 | 3.51 | 2.43 |
| (b) $n = 1000$ | Correct | Correct | $\pi$-*cov* | 0.00 | −0.1 | 1.17 | 0.79 |
| | | Incorrect | $\pi$-*cov* | −0.55 | −42.1 | 1.41 | 0.87 |
| | Incorrect | Correct | $\pi$-*cov* | 0.00 | −0.2 | 1.17 | 0.79 |
| | | Incorrect | $\pi$-*cov* | −1.49 | −100.6 | 2.10 | 1.56 |

In the spirit of Little and An (2004), let $S_i = S(\hat{\eta}_i)$ denote a vector of basis functions (e.g., a spline basis) that can serve to approximate the relationship between the mean of $y_i$ and $\hat{\eta}_i$, the estimated linear predictor from the $\pi$-model. Let $x_i^* = (x_i^T, S_i^T)^T$ denote the augmented vector of covariates, and let

$$\hat{\beta}^* = \left(\sum_i t_i x_i^* x_i^{*T}\right)^{-1} \left(\sum_i t_i x_i^* y_i\right) \quad (11)$$

denote the OLS-estimated coefficients from the augmented $y$-model. If $S_i$ is a spline basis of degree $k \geq 1$, the matrix inverse in (11) will not exist, because $S_i$ will contain a constant and a linear function of $\hat{\eta}_i$, which are themselves linear functions of $x_i$. Problems of collinearity can be alleviated by switching to a generalized inverse or by removing the offending terms from $S_i$; either approach leads to the same predicted values $\hat{m}_i^* = x_i^{*T}\hat{\beta}^*$. The propensity-covariate regression estimate for $\mu$ is then

$$\hat{\mu}_{\pi\text{-}cov} = \frac{1}{n}\sum_i \hat{m}_i^*. \tag{12}$$

TABLE 8
*Performance of inverse-propensity covariate regression estimators over 1000 samples from the artificial population*

| Sample size | $\pi$-model | $y$-model | Method | Bias | % Bias | RMSE | MAE |
|---|---|---|---|---|---|---|---|
| (a) $n = 200$ | Correct | Correct | $1/\pi$-*cov* | −0.09 | −3.7 | 2.48 | 1.69 |
| | | Incorrect | $1/\pi$-*cov* | 1.66 | 37.6 | 4.70 | 2.84 |
| | Incorrect | Correct | $1/\pi$-*cov* | 56.5 | 3.1 | 1804 | 1.76 |
| | | Incorrect | $1/\pi$-*cov* | −4236 | −3.4 | $1.3 \times 10^5$ | 7.79 |
| (b) $n = 1000$ | Correct | Correct | $1/\pi$-*cov* | 0.00 | −0.1 | 1.17 | 0.78 |
| | | Incorrect | $1/\pi$-*cov* | 0.59 | 31.3 | 1.97 | 1.26 |
| | Incorrect | Correct | $1/\pi$-*cov* | −50.4 | −3.2 | 1593 | 0.83 |
| | | Incorrect | $1/\pi$-*cov* | −7527 | −3.3 | $2.3 \times 10^5$ | 5.75 |

A particular case of (12) was proposed by Scharfstein, Rotnitzky and Robins (1999) who took $S_i = \hat{\pi}_i^{-1}$. Using the inverse-propensity as a single additional covariate is sufficient to achieve double robustness, because the estimate then becomes the solution to an AIPW estimating equation (Bang and Robins, 2005).

We tried (12) in our simulated example with a variety of spline bases: a quadratic spline with a single knot at the median of $\hat{\eta}_i$, a linear spline with knots at the quartiles, and so on. We found that these polynomial splines occasionally produced erratic predicted values of $y_i$ for nonrespondents with low propensities, driving up the variance of the regression estimate. The best performance was achieved by simply coarsening the $\hat{\eta}_i$'s into five categories and creating four dummy indicators to distinguish among them. In other words, we approximated the relationship between the mean of $y_i$ and $\hat{\pi}_i$ by a piecewise-constant function with discontinuities at the sample quintiles of $\hat{\pi}_i$. The performance of this estimate is summarized in Table 7. It performs better than any of the other DR methods when the $\pi$-model and $y$-model are both incorrect. It performs better than any method based on a $\pi$-model alone. Yet it is still inferior to the simple OLS regression estimate under the incorrect $y$-model.

In contrast, the regression estimate of Scharfstein, Rotnitzky and Robins (1999) that uses the inverse-propensity as a covariate behaves poorly under a misspecified $\pi$-model (Table 8). The performance of this method is disastrous when some of the estimated propensities are small.

## 4. DISCUSSION

Double robustness is an interesting theoretical property that can arise in many ways. It does not necessarily translate into good performance when neither model is correctly specified. Some DR estimators have been known to survey statisticians since the late 1970s. In survey contexts, these methods are not thought of as doubly robust but simply as robust, because the propensities are a known feature of the sample design. When propensities are unknown and must be estimated, care should be taken to select an estimator that is not overly sensitive to misspecification of the propensity model.

No single example can effectively represent all missing-data problems that researchers will see in practice. We constructed our simulation to vaguely resemble a quasi-experiment to measure the effect of dieting on body mass index (BMI) in a large sample of high-school students. The study has a pre-post design. Covariates $x_i$ measured at baseline include demographic variables, BMI, self-perceived weight and physical fitness, social acceptance and personality measures. The treatment $t_i$ is dieting (0 = yes, 1 = no) and the outcome $y_i$ is BMI one year later. The goal is to estimate an average causal effect of dieting among those who actually dieted. For that purpose, it suffices to treat the dieters as nonrespondents, set their BMI values to missing, and apply a missing-data method to estimate what the mean BMI for this group would have been had they not dieted. A simple linear regression of $y_i$ on $x_i$ among the nondieters yielded an $R^2$ of 0.81, just as in our simulated data, and boxplots of the linear predictors from a logistic propensity model looked similar to those of Figure 3(e). The large degree of overlap in the distributions of estimated propensities for the two groups makes a causal analysis seem feasible. The estimated propensities for some nondieters are very small. Keeping these nondieters in the analysis is highly desirable, because their covariate values closely resemble those of dieters; they provide excellent proxy information for predicting the missing values. Yet we found no obvious way to use these cases in an inverse-propensity weighted estimate, because they exerted too much influence.

Our simulation represents a situation where selection bias is moderate, good predictors of $y_i$ are available, both models are approximately but not exactly true, and some estimated propensities are nearly zero. In situations like these, a DR procedure that does not rely on inverse-propensity weighting may perform reasonably well, but there is no guarantee that it will outperform a procedure based solely on a $y$-model. A model that predicts $y_i$ reasonably well can enhance the performance of an approximate $\pi$-model. But in our simulations, we found no way to use the fitted propensities from the approximate $\pi$-model to reduce the bias from the approximate $y$-model. In every case, the bias correction applied to $\hat{\mu}_{OLS}$ tended to move the estimate in the wrong direction.

Some might argue that inference about $E(y_i)$ in the full population should not be attempted in a situation like the one shown in Figure 3(e), where the estimated propensities for a few nonrespondents fall outside the range of those seen among the respondents. In our opinion, requiring the empirical support of $P(\hat{\pi}_i \mid t_i = 1)$ to completely cover that of $P(\hat{\pi}_i \mid t_i = 0)$ is too stringent, especially given the sensitivity of these ranges to minor changes in the $\pi$-model. In all 1000 samples of $n = 200$ and $n = 1000$, the distributions of the estimated propensities overlapped sufficiently to compute a stratified estimate with five quintile-based groups. Moreover, although some of the estimated propensities were very small, none of the true propensities were actually zero, so the conditions required for DR estimation were not violated. Small propensities frequently occur when missing data are not missing by design, and data analysts need guidance on how to deal with them. One who would argue for the routine use of any kind of inverse-propensity weighted estimator ignores the obvious fact that these estimators cannot be used routinely.

Other evaluations of DR methods under dual misspecification have yielded mixed results, because the nature of the problems and degree of misspecification have varied. Davidian, Tsiatis and Leon (2005) presented a DR procedure analogous to $\hat{\mu}_{BC\text{-}OLS}$ for a pre–post analysis of a randomized clinical trial with dropout. Schafer and Kang (2005) evaluated their procedure and found that it performed slightly better than a parametric method based on multiple imputation (Rubin, 1987). In that analysis, each of the two treatment groups had its own $y$-model with $R^2$'s of about 0.5. Each group also had its own $\pi$-model, and the smallest fitted propensities were 0.14 and 0.07. It thus appears that the use of AIPW estimating equations can produce modest gains over a parametric method when the predictive power of the $y$-model is not too strong and the estimated propensities do not get close to zero.

The only other simulations that we know of that compare the performance of DR and non-DR methods under dual misspecification are those of Bang and Robins (2005). Their first example pertains to the estimation of a population mean $\mu$. They compare the performance of three methods—the unnormalized IPW estimate $n^{-1} \sum_i t_i \hat{\pi}_i^{-1} y_i$, the ordinary regression estimate $n^{-1} \sum_i x_i^T \hat{\beta}$, and the propensity-covariate estimate (12) that augments the $y$-model with $1/\hat{\pi}_i$—when the $\pi$- and $y$-models are correct and incorrect. In that example, the predictive ability of the correct $y$-model among the respondents is very strong ($R^2 = 0.94$), but the incorrect version is worthless ($R^2 = 0.001$); thus it maximally punishes the simple regression method when the $y$-model is misspecified. This approximates a situation where an analyst wants to impute missing values but has no idea how to use the covariates, so he simply ignores them and replaces all the missing values with $\bar{y}_i^{(1)}$. It may also represent a situation where all of the confounders are hidden from the analyst for purposes of $y$-modeling (though not, strangely, for purposes of $\pi$-modeling). Another noteworthy feature of that example is that all of the covariates in the propensity model have been dichotomized and the selection mechanism is weak; when $n$ is large, all of the estimated propensities fall nicely between 0.25 and 0.5. Bang and Robins (2005) present three additional examples to demonstrate the performance of DR estimates in more elaborate problems. In each case, all of the predictors in their $\pi$-models were dichotomized, which helps to keep the estimated propensities away from zero. Despite these features, none of their examples supports the claim that a DR method based on two incorrect models is clearly and simultaneously better than an IPW procedure based on an incorrect $\pi$-model and a simple imputation procedure based on an incorrect $y$-model. Only in the fourth example did the DR method outperform both of its competitors under dual misspecification, and even there the advantage was slight. Thus the results of Bang and Robins (2005) support our contention that two wrong models are not necessarily better than one.

## ACKNOWLEDGMENT

This research was supported by National Institute on Drug Abuse Grant P50-DA10075.